\documentclass{Interspeech2024}
\usepackage[export]{adjustbox} % 导入adjustbox包
\usepackage{subcaption}

\interspeechcameraready

\title{Exploring Sentence Type Effects on the Lombard Effect and Intelligibility Enhancement: A Comparative Study of Natural and Grid Sentences}

\name[affiliation={1}]{Hongyang}{Chen}
\name[affiliation={1,2,*}]{Yuhong}{Yang}
\name[affiliation={1}]{Zhongyuan}{Wang}
\name[affiliation={1,2}]{Weiping}{Tu}
\name[affiliation={1}]{Haojun}{Ai}
\name[affiliation={3}]{Song}{Lin}

\address{$^1$ NERCMS, School of Computer Science, Wuhan University\\
$^2$ Hubei Key Laboratory of Multimedia and Network Communication Engineering\\
$^3$ Guangdong OPPO Mobile Telecommunications Corp., China.}

\email{yangyuhong@whu.edu.cn}

\begin{document}

\maketitle

\newcommand\blfootnote[1]{%
\begingroup
\renewcommand\thefootnote{}\footnote{#1}%
\addtocounter{footnote}{-1}%
\endgroup
}

\begin{abstract}

This study explores how sentence types affect the Lombard effect and intelligibility enhancement, focusing on comparisons between natural and grid sentences. Using the Lombard Chinese-TIMIT (LCT) corpus and the Enhanced MAndarin Lombard Grid (EMALG) corpus, we analyze changes in phonetic and acoustic features across different noise levels. Our results show that grid sentences produce more pronounced Lombard effects than natural sentences. Then, we develop and test a normal-to-Lombard conversion model, trained separately on LCT and EMALG corpora. Through subjective and objective evaluations, natural sentences are superior in maintaining speech quality in intelligibility enhancement. In contrast, grid sentences could provide superior intelligibility due to the more pronounced Lombard effect. This study provides a valuable perspective on enhancing speech communication in noisy environments.

\end{abstract}
%

% \begin{keywords}
% Lombard effect, intelligibility enhancement, natural sentences 
% \end{keywords}

%
\section{Introduction}
\label{sec:intro}

% \textbf{test set objective evaluation}
% \textbf{the results of objective intel test in table 5 }
% \textbf{12 个有7个更好}
% \textbf{ 12 recording pairs}
% \textbf{400 recordings are converted by }

People involuntary increase their vocal effort to enhance speech intelligibility when speaking in noisy environments. This phenomenon is also known as the Lombard effect~\cite{lombard1911signe}. The changes in vocal effort involve not only loudness but also other phonetic and acoustic features, such as spectral tilt, fundamental frequency, vowel duration, the first and second formant frequencies, and so on~\cite{lombardgrid,mandarincxf}. Researchers also mimic the Lombard effect, called normal-to-Lombard conversion, to achieve intelligibility enhancement while ensuring naturalness~\cite{cyclegan,lidengshi2023}.

\blfootnote{$^\star$ Corresponding author. }
% \footnotetext[1]{Correspondence: yangyuhong@whu.edu.cn} 

\begin{figure}
    \centering

    \includegraphics[width=0.9\linewidth]{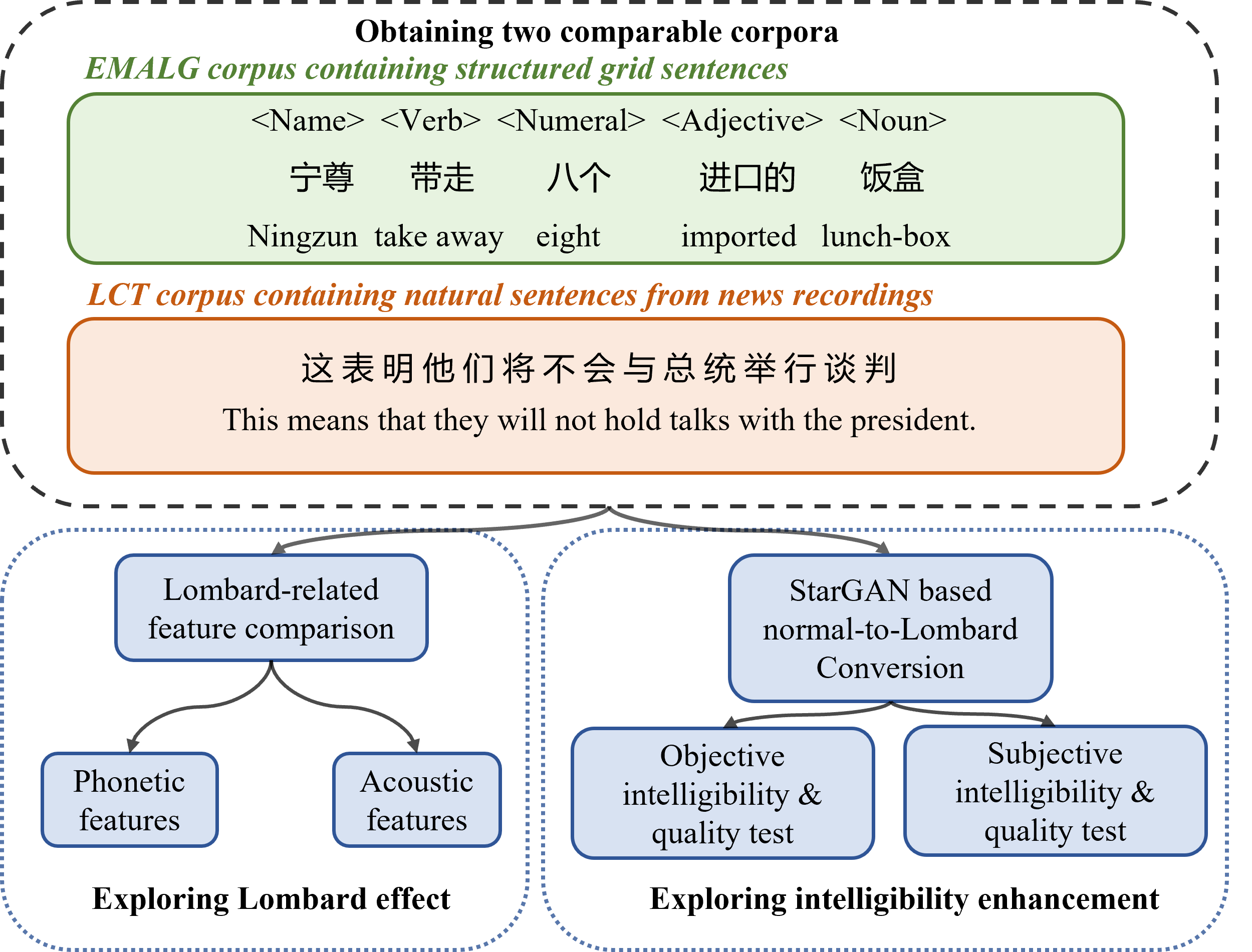}
    \caption{Block diagram of the comparative study.}
    \label{fig:motivation}
    \vspace*{-1.5em}

\end{figure}

To explore the Lombard effect, researchers have recorded many Lombard corpora (Table~\ref{tab:corpora}), varying in languages, sentence types, noise types, noise levels, etc. Natural sentences are common in everyday speech, so Lombard corpora based on natural sentences, such as UT-Scope~\cite{UT} and Emma Jokinen~\cite{finnish}, are better suited for real-world applications. However, since natural sentences have high-level linguistic cues and can predict upcoming words from previous ones, they are not suitable for speech perception~\cite{grid} and listening tests~\cite{gridpredict}, while grid sentences defined by a fixed syntactic structure and diverse word combinations (Figure~\ref{fig:motivation}) are unpredictable of the upcoming word. Therefore, some Lombard corpora based on grid sentences have been built, such as Lombard Grid~\cite{lombardgrid}, MALG~\cite{mandarincxf}, and EMALG~\cite{lbf}. Previous studies have shown that the Lombard effect is sensitive to the acoustic context, including languages~\cite{mandarincxf}, noise types~\cite{competingspeaker}, and noise levels~\cite{UT}, but they have overlooked sentence types on the Lombard effect. Since the current Lombard corpora also vary in terms of languages, noise levels, and noise types, they are insufficient to investigate the impact of sentence types on the Lombard effect. We need to obtain two comparable Lombard corpora with the same setups except sentence type to explore the impact of sentence types on the Lombard effect. 

Normal-to-Lombard conversions~\cite{cyclegan,ligangstargan,lidengshi2023} are based on Lombard corpora~\cite{finnish,german,lombardgrid} with different sentence types to mimic the Lombard effect for intelligibility enhancement. For instance, CycleGAN~\cite{cyclegan} and D2StarGAN~\cite{lidengshi2023} are based on natural sentences Lombard corpora~\cite{finnish,german}, StarGAN~\cite{ligangstargan} is based on grid sentences Lombard corpora~\cite{lombardgrid}. We are interested in exploring the impact of different sentence types on intelligibility enhancement.

\begin{table*}[]
\caption{Lombard corpus settings overview. Initial noise levels represent quiet backgrounds, not triggering the Lombard effect. "SSN" refers to speech-shaped noise.}
% The first item in each row under the Noise level column represents the noise level of normal speech.
% The first item in each row of Noise level corresponds to the noise level of normal speech (speaking in quiet environment without Lombard effect).

\centering
\label{tab:corpora}
% \resizebox{\textwidth}{!}
% {
\begin{tabular}{lllll}
\hline
Name                                                  & Sentence type & Language & Noise type            & Noise level (dB)             \\ \hline
Lombard Grid~\cite{lombardgrid} & grid          & English  & SSN                   & 30/80                        \\
Cooke~\cite{competingspeaker}   & grid          & English  & competing speaker/SSN & quiet/82/89/96               \\
UT-Scope~\cite{UT}              & natural       & English  & pink/car/crowd        & quiet/ (65/75/85) for pink   \\
                                                      &               &          &                       & (70/80/90) for car and crowd \\
Emma Jokinen~\cite{finnish}     & natural       & Finnish  & car/pub               & quiet/65/80                  \\
M. Sołoducha~\cite{german}      & natural       & German   & babble                & quiet/55/70                  \\
MALG~\cite{mandarincxf}                 & grid          & Mandarin & SSN                   & 30/55/70/80                     \\
EMALG~\cite{lbf}                & grid          & Mandarin & SSN                   & 40/55/80                     \\
LCT (ours)                                            & natural       & Mandarin & SSN                   & 40/55/80                     \\ \hline
\end{tabular}
% }、
    \vspace*{-0.5em}
\end{table*}

An overview diagram of our approach is provided in Figure~\ref{fig:motivation}. In this study, we record the Lombard Chinese-TIMIT (LCT) corpus using natural sentences and compare it with EMALG corpus, which employs grid sentences. The comparative study includes the following two contributions:

\begin{itemize}
    \item We explore how sentence types affect the Lombard effect by examining phonetic and acoustic feature changes. Results indicate that while both types show similar trends in feature changes, grid sentences exhibit a more pronounced Lombard effect than natural sentences. 
    \item To explore the impact of different sentence types on intelligibility enhancement, we develop and evaluate two Normal-to-Lombard conversion models based on the StarGAN architecture, trained respectively on the EMALG and LCT corpus. These models convert normal speech into Lombard speech at 80 dBA. Subjective and objective evaluations show grid sentences enhance intelligibility more effectively, whereas natural sentences maintain better speech quality.
\end{itemize}
% 1. 
% 2. 

% \section{Lombard effect comparison}
\section{Exploring Lombard effect}

\subsection{Obtaining two comparable Lombard corpora}
We record the Lombard Chinese-TIMIT (LCT) corpus to examine the impact of natural sentences on the Lombard effect. The selection criteria for participants, recording environment, noise levels, noise type, equipment calibration, and recording setups are aligned with those used in the EMALG corpus, except for the sentence materials. We use speech-shaped noise (SSN,~\cite{ssn}) to record normal speech at a noise level of 40 dBA (normal speech style, N) and Lombard speech at noise levels of 55 dBA (the first Lombard speech style, L1) and 80 dBA (the second Lombard speech style, L2).

EMALG~\cite{lbf} uses 50 words covering all phonemes to compose Mandarin grid sentences, and we choose Chinese TIMIT~\cite{chinesetimit} to get natural Mandarin sentences. All sentences are 10-20 characters long and selected from the corpus of Chinese Gigaword Fifth Edition~\cite{gigaword}, which is a comprehensive archive of newswire text data from Chinese news sources. In Chinese TIMIT, there are three groups of sentences: "Calibration", "Shared" and "Unique". For each speaker, we prepared a total of 100 sentences, consisting of 20 Calibration sentences, 40 Shared sentences, and the first 40 sentences from 60 Unique sentences. EMALG~\cite{lbf} recruited 34 students at Wuhan University to read the sentences, and we recruited 36 students (18 males and 18 females) at university to read the sentences. All of them speak Standard Chinese, achieving Class 2 Level 1 or better on the national standard Mandarin proficiency test. All participants are paid for their contributions.

% We use all sentences of 20 Calibration sentences, 40 Shared sentences, and the first 40 sentences of 60 Unique sentences, a total of 100 sentences for each speaker.

\subsection{Lombard-related features}
\label{sec:analysis}

% 七个声学特征我们提取了四个声学相关的特整
We extract a total of seven Lombard-related features from the EMALG and LCT corpora. Four of these are phonetic features, including average vowel duration (vowel duration), the ratio of total vowel to utterance duration (vowel ratio), and the average first and second formant frequencies for vowels (F1 and F2). Additionally, there are three acoustic features, including fundamental frequency on a semitone frequency scale (F0), an estimate of perceived signal intensity from an auditory spectrum (loudness~\cite{loudness}) and an average logarithmic energy ratio between 50-1000 Hz and 1-15 kHz (alpha ratio~\cite{alpha}).

Four phonetic features are extracted by Montreal Forced Aligner (MFA,~\cite{mfa}) and Praat~\cite{praat}. We use MFA with both EMALG and LCT to train an alignment model and generate \emph{TextGrid}. From \emph{TextGrid}, we calculate the vowel duration and vowel ratio. Then, we employ Praat to estimate the F1 and F2. The openSMILE~\cite{opensmile} tools are used to extract three acoustic features (F0, loudness and alpha ratio) from the Geneva Minimalistic Acoustic Parameter Set~\cite{gemaps}.

As in Figure~\ref{fig:motivation}, we focus on the feature differences between N and L1, as well as between N and L2, to examine the Lombard effect at low and high noise levels, respectively. Paired-sample t-tests are employed to determine the significant difference between two speaking styles across genders (male (M) speakers, female (F) speakers and all (ALL) speakers). We calculate the mean and standard deviation of seven Lombard-related features in three noise conditions (N, L1, and L2) across genders. We also calculate the increase of Lombard-related features between different styles in EMALG and LCT.

\section{Exploring intelligibility enhancement}

% \subsection{StarGAN}
\subsection{Comparable StarGAN models}

\begin{figure}
    \vspace*{-1em}
    \centering
    \includegraphics[width=1\linewidth]{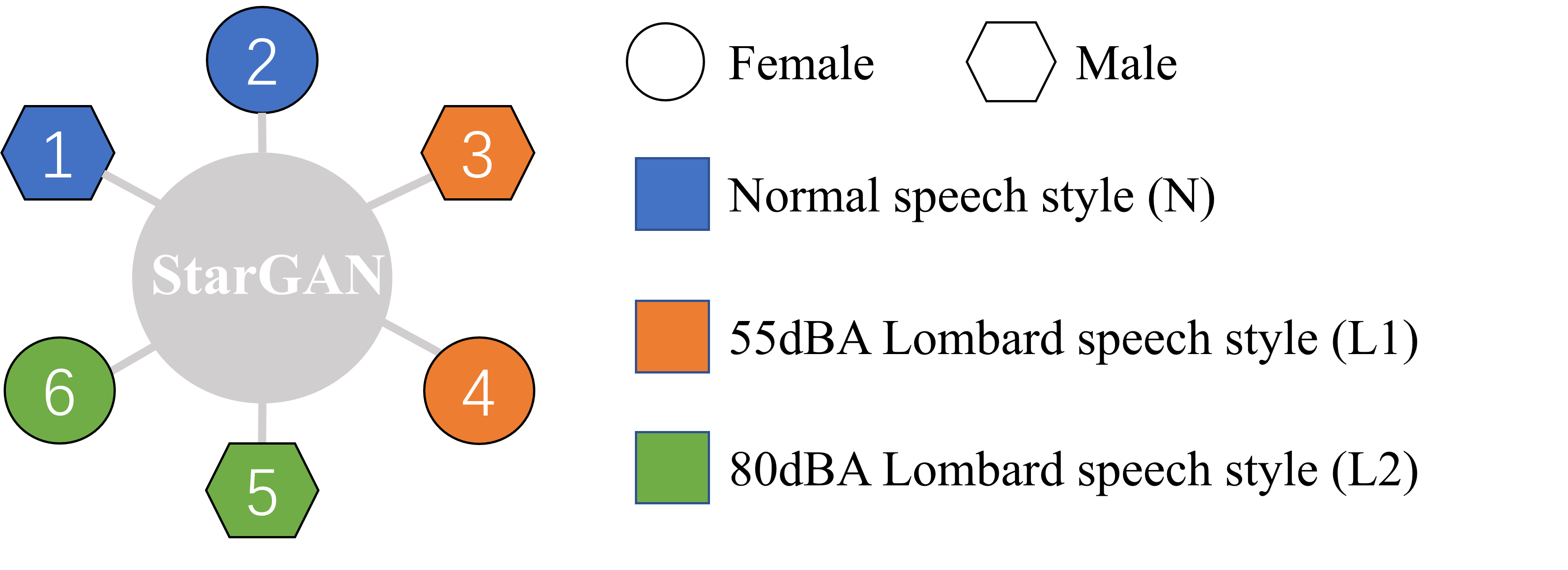}
    \caption{Our implemented StarGAN architecture with six domains for intelligibility enhancement. In inference, normal speech are converted to L2 speech.} 
    \label{fig:stargan}
    \vspace*{-2em}
\end{figure}

The recent method~\cite{ligangstargan,lidengshi2022,lidengshi2023} for normal-to-Lombard conversion is still based on StarGAN. Therefore, we opted for StarGAN~\cite{ligangstargan} to implement intelligibility enhancement. Given the differences in the Lombard effect between males and females~\cite{lbf}, Figure~\ref{fig:stargan} presents the StarGAN architecture with six domains: three speaking styles (N, L1, L2) and two genders (female and male). The network structure and parameter settings are implemented according to~\cite{ligangstargan}. Recognizing the prevalence of natural sentences in daily communication, we randomly choose 4 LCT speakers (2 female, 2 male) as the test set, with the remaining 32 speakers forming the training set. We developed EMALG-StarGAN, trained on 34 EMALG speakers, and LCT-StarGAN, using 32 LCT speakers (16 female, 16 male). Audio is downsampled from 48 kHz to 16 kHz. Aiming to boost intelligibility, L2 is the target style (5\textsuperscript{th} and 6\textsuperscript{th} domain) in the inference phase, converting normal speech from the source domains (1\textsuperscript{st} and 2\textsuperscript{nd} domain) to L2.

\subsection{Objective evaluation comparison}

The 400 recordings of normal speech in the test set are converted by LCT-StarGAN and EMALG-StarGAN, resulting in 400 pairs of test sequences of target Lombard style (L2).

\textit{Objective intelligibility} is measured using the Gaussian variant of Speech Intelligibility in Bits (SIIB,~\cite{siib}) and an Automatic Speech Recognition system (ASR,~\cite{whisper}). SIIB is based on the mutual information between a clean reference and a noisy signal. We employ the word recognition rate (WRR,~\cite{wrr}) as the measure of ASR results. We apply SSN to LCT-StarGAN speech and EMALG-StarGAN speech at Signal-to-Noise Ratio (SNR) levels of -8.5, -6, and -3.5 to obtain noisy signals (covering 50\% speech intelligibility~\cite{srt}).

\textit{Objective quality} is evaluated using MOSNET~\cite{mosnet} and UTMOS~\cite{utmos} for clean LCT-StarGAN speech and clean EMALG-StarGAN speech. MOSNET and UTMOS are two deep learning-based non-intrusive objective metrics for speech quality. All converted speech is normalized to have the same energy.

\subsection{Subjective evaluation comparison}
The subjective evaluation includes an intelligibility test and a quality test. Eighteen students at university, aged 18 to 26, participate in the listening tests as listeners. Sounds are played to the listeners in a quiet room using HD 300 pro headphones. We select 12 recordings of normal speech from the test set (3 sentences randomly chosen from the Unique sentences of 4 speakers), which are then converted using the LCT-StarGAN and EMALG-StarGAN models, resulting in 12 pairs of test sequences of target Lombard style (L2). The two tests take approximately one hour to complete for the subjects.

\textit{Subjective intelligibility test} is conducted to clarify the performance of EMALG-StarGAN and LCT-StarGAN. We establish 3 SNR levels at -8.5, -6 and -3.5. In order to avoid the memorization effect, the 12 pairs of test sequences are allocated to the three SNR levels in the same proportion, resulting in 4 pairs of test sequences per SNR. To reduce individual differences, 18 students from university, of whom 9 listen to the 12 EMALG-StarGAN speech and 9 listen to the 12 LCT-StarGAN speech. We employ WRR as the measure of intelligibility. Testers are required to adjust the volume to their maximum acceptable level before starting. Testers will then repeatedly listen to the played speech and write the \emph{pinyin} of the words they hear in the input box. Testers are allowed to guess words they are not sure about.

\textit{Subjective quality test} is conducted via MUSHRA (MUltiple Stimuli with Hidden Reference and Anchor~\cite{mushra})) test\footnote{https://github.com/audiolabs/webMUSHRA}. Each trial aims to evaluate the quality of the converted utterances. In a single trial, the listeners are presented with an 80dBA Lombard-style reference sample and two unmarked samples, each converted from normal speech by EMALG-StarGAN and LCT-StarGAN. The listeners are required to score these samples using a rating scale ranging from 0 to 100. In the end, every test sample receive at least 12 subjective quality scores. Participants perform the subjective quality test only after completing the subjective intelligibility test to prevent memorization effects.

% To avoid the memorization effect on the subjective Intelligibility Test, participants performed the subjective quality test only after completing the subjective Intelligibility Test. 

\section{Result and analysis}

\subsection{Lombard-related features comparison}

\begin{figure}[h]
    \centering
    \includegraphics[width=0.9\linewidth]{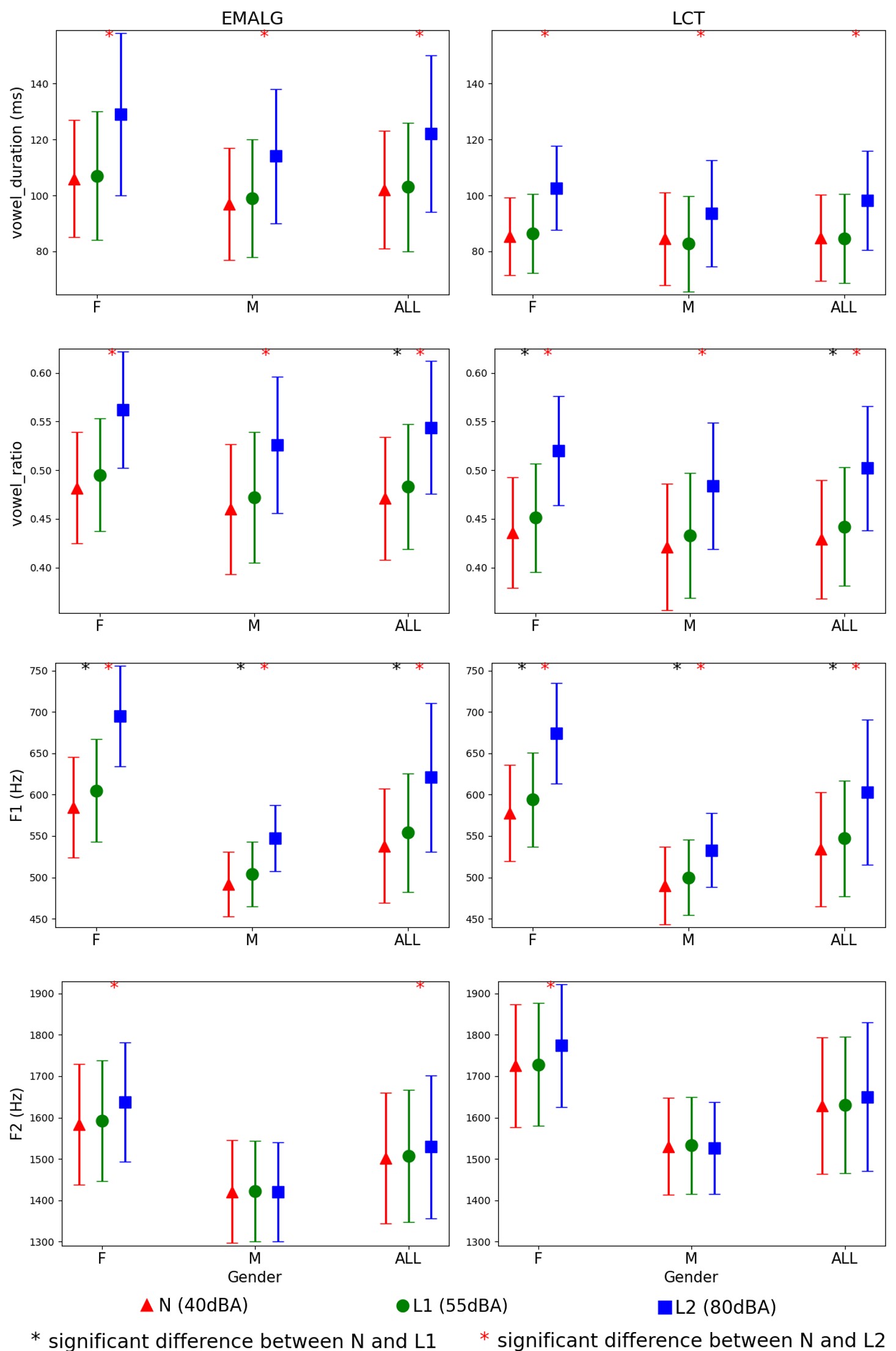}

\caption{Lombard effect analysis 1. The mean and standard deviation of vowel duration, vowel ratio, F1, and F2 of male (M) speakers, female (F) speakers and all (ALL) speakers. Error bars represent standard deviation. * means a significant difference (p $<$ 0.001) in t-test between two different speaking styles across genders.}
% Natural represents LCT, and Grid represents EMALG. 
    \label{fig:vowel}
    \vspace*{-1em}

\end{figure}

\begin{figure}
    \centering
    \includegraphics[width=0.9\linewidth]{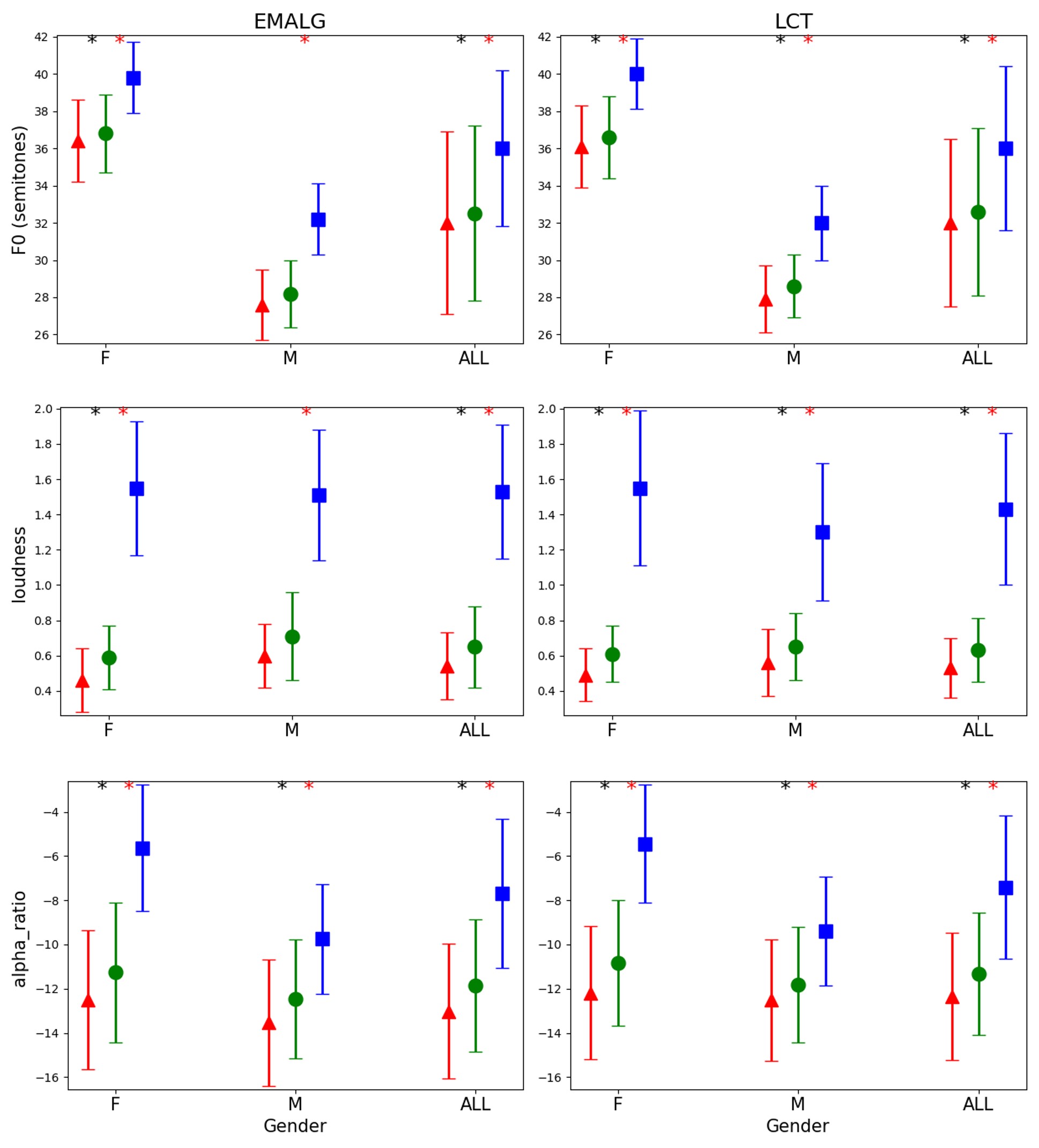}
    \caption{Lombard effect analysis 2. The mean and standard deviation of F0, loudness, and alpha ratio.}
    \label{fig:formant}
    \vspace*{-1.2em}

\end{figure}

The results of phonetic and acoustic features are shown in Figure~\ref{fig:vowel} and Figure~\ref{fig:formant}. As the noise level increases, we find that both sentence types exhibit generally similar trends in feature variations. We also have made some interesting observations:

Between N and L1, the Lombard effect is more pronounced in grid sentences. Between N and L1, grid sentences exhibit slightly greater improvements in the F1, loudness, and alpha ratio compared to natural sentences. For all speakers, the F1 has increased by 2.4\% (LCT) and 2.8\% (EMALG). The loudness has increased by 18.9\% (LCT) and 20.4\% (EMALG). The alpha ratio has increased by 1.02 (LCT) and 1.15 (EMALG) in the logarithm domain. However, natural sentences show more gender-dependent significant differences. Specifically, there's a significant difference in vowel ratio for females in LCT but not in EMALG. Similarly, there are significant differences in F0 and loudness for males in LCT but not in EMALG. These indicate that the Lombard effect of natural sentences is easier to elicit across different genders at low noise levels.

% \vspace{1pm}

Similarly, between N and L2, grid sentences exhibit greater improvements in the vowel duration, F1, loudness, and alpha ratio compared to natural sentences. For all speakers, the vowel duration has increased by 15.7\% (LCT) and 18.6\% (EMALG). The F1 has increased by 12.9\% (LCT) and 15.2\% (EMALG). The loudness has increased by 170.8\% (LCT) and 183.3\% (EMALG). The alpha ratio has increased by 4.94 (LCT) and 5.34 (EMALG). These indicate that grid sentences with a fixed syntactic structure have fuller vowel articulation, exhibit greater changes in tongue downward position for vowels, experience more enhancement in energy, and demonstrate a greater increase in the movement of low-frequency energy towards high frequencies. However, compared to grid sentences, natural sentences exhibit a greater increase in vowel ratio (17.0\% for LCT and 14.0\% for EMALG). The greater increase in vowel ratio is attributed to the memorization effect of natural sentences shortening the duration of the sentences.

As the noise level increases, the changes in F0 and F2 are unaffected by sentence types. The F0 values for both sentence types remain nearly the same across all noise levels and genders. Both natural and grid sentences exhibit significant differences in F2 for female speakers between N and L2, with increases of 2.8\% (LCT) and 3.4\% (EMALG). We believe the slight differences don't reflect the influence of sentence types on the variation of F2. Thus, as the noise level increases, sentence types don't affect the changes in fundamental frequency and tongue forward position for vowels.

% \subsection{Intelligibility enhancement application comparison}
\subsection{StarGAN models comparison}
% \subsubsection{Objective evaluation comparison}

The results of the objective intelligibility test in Table~\ref{tab:objective} show that EMALG-StarGAN speech has higher intelligibility than LCT-StarGAN speech, with a 4\% increase in SIIB and an 18\% increase in ASR. This may be attributed to the more pronounced Lombard effect observed in grid sentences.

The results of the objective quality test in Table~\ref{tab:objective} show that LCT-StarGAN speech exhibits superior quality compared to EMALG-StarGAN speech, with a 1.2\% increase in UTMOS and a 2.5\% increase in MOSNET. This indicates that natural sentences can enable the normal-to-Lombard conversion model to maintain speech quality better.

\begin{table}[]
\centering
\caption{Objective evaluation results for LCT-StarGAN and EMALG-StarGAN. Higher is better. The intelligibility results are average across three SNR conditions. WRR is used to measure the results of ASR.}
%   可懂度结果是三个SNR下的平均结果Average objective scores of under 3 SNR
\label{tab:objective}
\resizebox{0.98\linewidth}{!}{
\begin{tabular}{c|cccc}
\hline
              & \multicolumn{2}{c}{Intelligibility} & \multicolumn{2}{c}{quality} \\ \hline
metrics       & SIIB             & ASR (\%)             & UTMOS        & MOSNET       \\ \hline
LCT-StarGAN & 21.253           & 13.601           & \textbf{2.684}        & \textbf{3.260}        \\
EMALG-StarGAN  & \textbf{21.963}           & \textbf{16.145}           & 2.651        & 3.181        \\ \hline
\end{tabular}}
    \vspace*{-0.3em}
\end{table}

\begin{figure}
  \centering
    \includegraphics[width=0.9\linewidth]{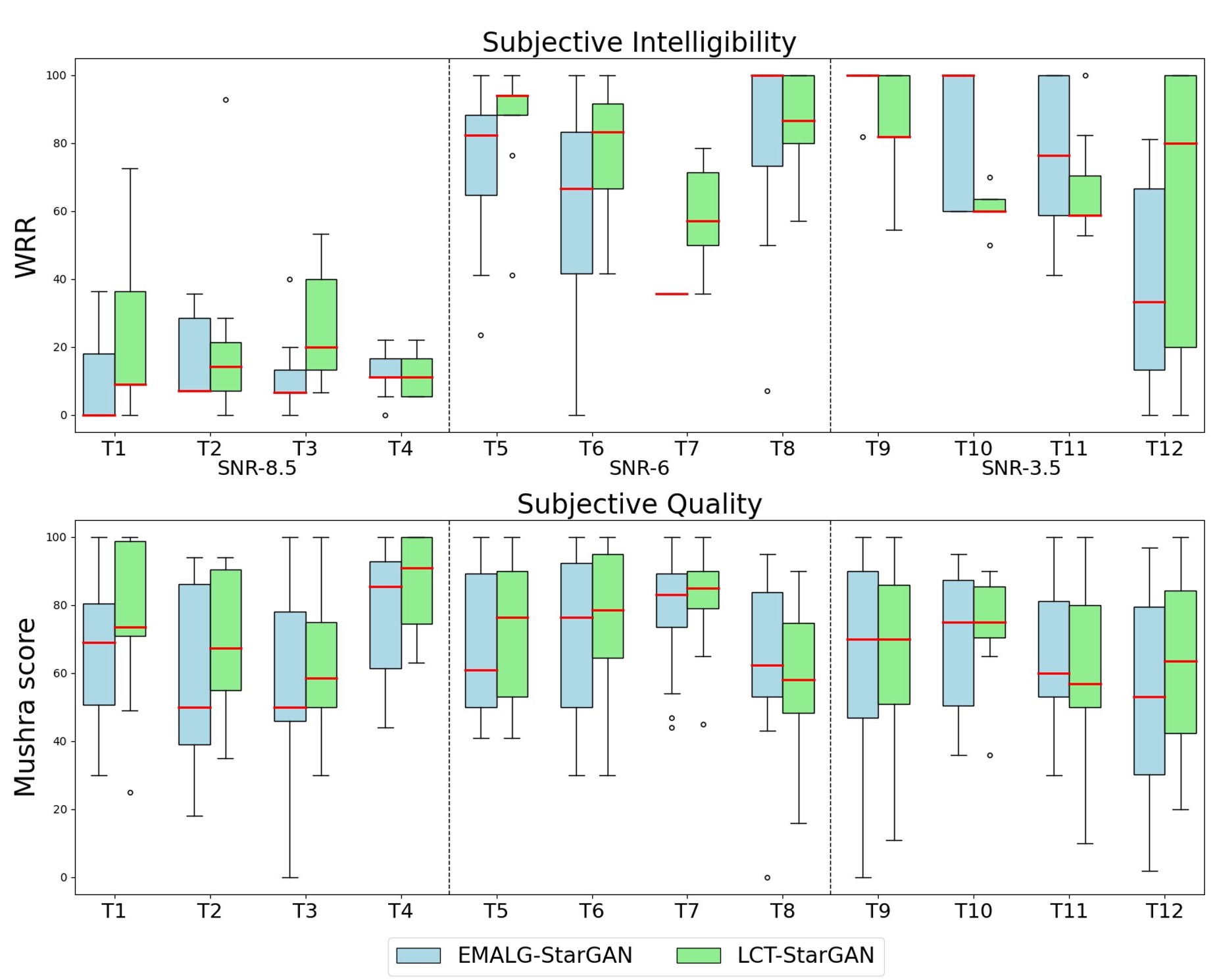}
    % \caption{}
    % \caption{}
  \caption{Boxplots of subjective evaluation results for LCT-StarGAN and EMALG-StarGAN. Higher is better. T1-T12 correspond to 12 pairs of test sequence.}
  \label{fig:total}
  \vspace*{-1em}

\end{figure}

% \subsubsection{Subjective evaluation comparison}
The results of the subjective intelligibility test in Figure~\ref{fig:total} show that LCT-StarGAN speech performs better than EMALG-StarGAN speech in 7 out of 12 recording pairs, which is not consistent with the results of the objective intelligibility test. Due to the limited number of recording pairs covered in our study, sample bias may lead to the inconsistency. We believe that grid sentences could provide superior intelligibility.

The results of the subjective quality test in Figure~\ref{fig:total} show that LCT-StarGAN speech performs better than EMALG-StarGAN speech in 8 out of 12 recording pairs. The average MUSHRA score of LCT-StarGAN speech is higher than EMALG-StarGAN, surpassing 6\%. This is consistent with the results of the two objective quality metrics.

% \begin{figure}
%   \centering
%   \begin{subfigure}{\linewidth}
%     \includegraphics[width=\linewidth]{subjective_intelligibility.jpg}
%     % \caption{}
%     \label{fig:subjective_intelligibility}
%   \end{subfigure}
%   \begin{subfigure}{\linewidth}
%     \includegraphics[width=\linewidth]{subjective_quality.jpg}
%     % \caption{}
%     \label{fig:subjective_quality}
%   \end{subfigure}
%   \caption{Boxplots of subjective evaluation results for LCT-StarGAN and EMALG-StarGAN. T1-T12 correspond to 12 pairs of test sequence.}
%   \label{fig:total}
% \end{figure}

\section{Conclusion}

In this work, we conduct a comparative study of the Lombard effect and intelligibility enhancement between natural and grid sentences. Through Lombard-related features analysis, we find that both sentence types exhibit generally similar trends in feature variations. However, grid sentences, which have a fixed syntactic structure, exhibit a more pronounced Lombard effect, particularly in greater changes of vowel duration, loudness, spectral tilt, and tongue downward position for vowels. Changes in fundamental frequency and tongue forward position for vowels are not affected by sentence types. Through subjective and objective tests conducted on normal-to-Lombard conversion model trained on two Lombard corpora, we find that natural sentences could maintain better quality, whereas grid sentences could provide superior intelligibility. These findings emphasize the important role of sentence types in the Lombard effect and intelligibility enhancement, providing valuable insights for enhancing speech communication in noisy environments. In the future, we will increase the sample size of subjective tests and explore more normal-to-Lombard conversion models.

\noindent \textbf{Acknowledgments.}
This research is funded in part by the National Natural Science Foundation of China (62171326, 62071342), Key Research and Development Program of Hubei Province (220171406) and Guangdong OPPO Mobile Telecommunications Corp.

\bibliographystyle{IEEEtran}
\bibliography{mybib}

\end{document}